# Spinless particles in the field of unequal Scalar-Vector Yukawa potentials

M. Hamzavi[1*], S. M. Ikhdair[2 +], K.-E. Thylwe[3 †]

[1]*Department of Basic Sciences, Shahrood Branch, Islamic Azad University, Shahrood, Iran*

[2]*Physics Department, Near East University, 922022 Nicosia, North Cyprus, Mersin 10, Turkey*

*and*

[2]*Physics Department, Faculty of Science, An-Najah National University, Nablus, West Bank, Palestine*

[3]*KTH-Mechanics, Royal Institute of Technology, S-100 44 Stockholm, Sweden*

[*]*Corresponding author: Tel.:+98 273 3395270, fax: +98 273 3395270*

[*]Email: majid.hamzavi@gmail.com

[+] Email: sikhdair@neu.edu.tr; sikhdair@gmail.com

[†] Email: ket@mech.kth.se

**Abstract**

We present analytical bound state solutions of the spin-zero Klein-Gordon (KG) particles in the field of unequal mixture of scalar and vector Yukawa potentials within the framework of the approximation scheme to the centrifugal potential term for any arbitrary $l$-state. The approximate energy eigenvalues and unnormalized wave functions are obtained in closed forms using a simple shortcut of the Nikiforov-Uvarov (NU) method. Further, we solve the KG-Yukawa problem for its exact numerical energy eigenvalues via amplitude phase (AP) method to test the accuracy of the present solutions found by using the NU method. Our numerical tests using energy calculations demonstrate the existence of inter-dimensional degeneracy amongst energy states of the KG-Yukawa problem. The dependence of the energy on the dimension $D$ is numerically discussed for spatial dimensions $D = 2 - 6$.

## 1. Introduction

The Yukawa potential or static screening Coulomb (SSC) potential is often used to compute bound-state normalizations and energy levels of neutral atoms [1-6] which have been studied over the past years. It is known that SSC potential yields reasonable results only for the innermost states when $Z$ is large. However, for the outermost and middle atomic states, it gives rather poor results. The bound-state energies for the SSC potential with $Z=1$ have been studied in the light of the shifted large-$N$ expansion method [7]. For example, Chakrabarti and Das presented a perturbative solution of the Riccati equation leading to analytical superpotential for the Yukawa potential [8]. Ikhdair and Sever investigated energy levels of neutral atoms by applying an alternative perturbative scheme in solving the Schrödinger equation for the Yukawa potential model with a modified screening parameter [9]. They also studied bound states of the Hellmann potential, which represents the superposition of the attractive Coulomb potential $-a/r$ and the Yukawa potential $b\exp(-\delta r)/r$ of arbitrary strength $b$ and screening parameter $\delta$ [10]. Some authors studied relativistic and non-relativistic equations with different potentials [11-44]. The aim of the present work is to investigate the KG equation in arbitrary dimension $D$ [45] with unequal mixture of scalar and vector Yukawa potentials:

$$V(r)=-V_0\frac{e^{-ar}}{r}, \tag{1a}$$

$$S(r)=-S_0\frac{e^{-ar}}{r} \tag{1b}$$

$$S(r)=\beta V(r), \quad -1\leq\beta\leq 1 \tag{1c}$$

where $V_0=\alpha Z$ , $\alpha=(137.037)^{-1}$ is the fine-structure constant and $Z$ is the atomic number and $a$ is the screening parameter [1]. Also, $\beta$ is an arbitrary a constant demonstrating the ratio of scalar potential to the vector potential [46]. When $\beta=1$, we have equal mixture, i.e., $S(r)=V(r)$, representing the exact spin symmetric limit $\Delta(r)=S(r)-V(r)=0$ (the potential difference is exactly zero). On the other hand, when $\beta=-1$, we have $S(r)=-V(r)$, representing the exact pseudospin (p-spin) symmetric limit $\Sigma(r)=S(r)+V(r)=0$ (the potential sum is exactly zero). The strong singular centrifugal term is being approximated within the framework of an improved approximation scheme. Further, the spinless $D$-dimensional KG equation with the scalar and vector Yukawa potentials is solved using the parametric generalization of the NU method [47-49] to obtain approximate analytical energy eigenvalues and corresponding wave functions for any $l$-state. The approximated numerical energy eigenvalues in the present model are compared with exact numerical results obtained by AP method [50-53].

The present work is organized as follows. In section 2, the generalized parametric NU and AP methods are briefly introduced. In section 3, we give a review to the KG equation in $D$-dimensional space and then obtain the bound state solutions of the hyperradial KG equation with unequal mixture of scalar and vector Yukawa potentials by using a shortcut of the NU method. The exact and approximate numerical results are also given by the AP and NU methods, respectively. Finally, we end with our concluding remarks in section 4.

## 2. Methods of Analysis

In this section, we give a brief review to the analytical NU and the numerical AP methods.

### 2.1. Parametric NU Method

This powerful mathematical tool solves second-order differential equations. It starts by considering the following differential equation [47-49]

$$\psi_n''(s)+\frac{\tilde{\tau}(s)}{\sigma(s)}\psi_n'(s)+\frac{\tilde{\sigma}(s)}{\sigma^2(s)}\psi_n(s)=0, \tag{2}$$

where $\sigma(s)$ and $\tilde{\sigma}(s)$ are polynomials, at most of second degree, and $\tilde{\tau}(s)$ is a first-degree polynomial. To make the application of the NU method simpler and straightforward without need to check the validity of solution, we present a shortcut for the method. At first we write the general form of the Schrödinger-like equation (2) in a more general form:

$$\psi_n''(s)+\left(\frac{c_1-c_2s}{s(1-c_3s)}\right)\psi_n'(s)+\left(\frac{-p_2s^2+p_1s-p_0}{s^2(1-c_3s)^2}\right)\psi_n(s)=0, \tag{3}$$

satisfying the wave functions:

$$\psi_n(s)=\phi(s)y_n(s). \tag{4}$$

Now, comparing Eq. (3) with its counterpart Eq. (2), one can obtain the following identifications:

$$\tilde{\tau}(s)=c_1-c_2s,\quad \sigma(s)=s(1-c_3s),\quad \tilde{\sigma}(s)=-p_2s^2+p_1s-p_0, \tag{5}$$

Further, following the NU method [47], we can obtain the bound-state energy equation [48,49]

$$c_2n-(2n+1)c_5+(2n+1)\left(\sqrt{c_9}+c_3\sqrt{c_8}\right)+n(n-1)c_3+c_7+2c_3c_8+2\sqrt{c_8c_9}=0. \tag{6}$$

We also find the functions

$$\rho(s) = s^{c_{10}}(1-c_3 s)^{c_{11}}, \quad \phi(s) = s^{c_{12}}(1-c_3 s)^{c_{13}}, \; c_{12} > 0, \; c_{13} > 0,$$

$$y_n(s) = P_n^{(c_{10}, c_{11})}(1-2c_3 s), \; c_{10} > -1, \; c_{11} > -1, \tag{7a}$$

which are necessary in calculating the wave functions

$$\psi_{nl}(s) = N_{nl} s^{c_{12}}(1-c_3 s)^{c_{13}} P_n^{(c_{10}, c_{11})}(1-2c_3 s), \tag{7b}$$

where $P_n^{(\mu,\nu)}(x)$, $\mu > -1$, $\nu > -1$, $x \in [-1,1]$ are Jacobi polynomials with constant parameters [30]

$$c_4 = \frac{1}{2}(1-c_1), \qquad c_5 = \frac{1}{2}(c_2 - 2c_3),$$

$$c_6 = c_5^2 + p_2; \qquad c_7 = 2c_4 c_5 - p_1,$$

$$c_8 = c_4^2 + p_0, \qquad c_9 = c_3(c_7 + c_3 c_8) + c_6,$$

$$c_{10} = c_1 + 2c_4 + 2\sqrt{c_8} - 1 > -1, \qquad c_{11} = 1 - c_1 - 2c_4 + \frac{2}{c_3}\sqrt{c_9} > -1, \; c_3 \neq 0,$$

$$c_{12} = c_4 + \sqrt{c_8} > 0, \qquad c_{13} = -c_4 + \frac{1}{c_3}(\sqrt{c_9} - c_5) > 0, \; c_3 \neq 0, \tag{8}$$

where $c_{12} > 0$, $c_{13} > 0$ and $s \in [0, 1/c_3]$, $c_3 \neq 0$.

When considering a more special case of $c_3 = 0$, the wave function (7b) becomes

$$\lim_{c_3 \to 0} P_n^{(c_{10}, c_{11})}(1-2c_3 s) = L_n^{c_{10}}(c_{11} s), \; \lim_{c_3 \to 0}(1-c_3 s)^{c_{13}} = e^{c_{13} s},$$

$$\psi(s) = N s^{c_{12}} e^{c_{13} s} L_n^{c_{10}}(c_{11} s). \tag{9}$$

where $L_n^{\alpha}(x)$ are the associated Laguerre polynomials.

### 2.2. Amplitude-Phase Method

The amplitude-phase (AP) method used for calculating bound states was presented by Korsch and Laurent in 1981 [50]. The Schrödinger equation can be converted into an equation for the so-called amplitude function which is a formal relation to a local wave number that determines the so-called phase function. It begins by writing the radial solution $R_{nl}(r)$ of the Schrödinger equation in the form [50-53]:

$$R_{nl}(r) = u(r) \sin \phi(r), \tag{10}$$

with the imposed relation

$$\phi'(r) = u(r)^{-2}, \tag{11}$$

one derives a non-linear second-order differential equation for the amplitude function

$$\frac{d^2u(r)}{dr^2} + \left\{ 2[E_{nl} - V(r)] - \frac{l(l+1)}{r} \right\} u(r) = \frac{1}{u(r)^3}. \tag{12}$$

Equation (12) is the so called a Milne-type equation [50]. A phase reference condition (because of an arbitrary integration constant) is required, and it is formally accomplished with

$$\phi(r) \to 0, \qquad \text{as } r \to 0. \tag{13}$$

Integration of (12) goes in two steps; from the potential minimum towards $+0$, and from the potential minimum towards $+\infty$ using initially the (first-order) semiclassical formula for the amplitude given as follows (cf. Ref. [50]):

$$u(r_{\min}) \approx \left[ 2(E_{nl} - V(r_{\min})) - \frac{l(l+1)}{(r_{\min})^2} \right]^{-1/4}, \;\; u'(r_{\min}) = 0. \tag{14}$$

The condition for the wave function to vanish at $r \to +\infty$ then is

$$\phi(+\infty) = (1+n)\pi, \;\; n = 0, 1, ..., \tag{15}$$

where $n$ is the (nodal) radial quantum number. The quantization condition (15) is solved by using a Newton's iteration with respect to the energy. Further details of the numerical aspects are recently given in [53].

## 3. Hyperradial Part of the KG Equation in $D$ -Dimensional Space

In spherical coordinates, the KG equation with vector $V(r)$ and scalar $S(r)$ potentials can be written as (in units of $\hbar = c = 1$) [46]

$$\left[ \Delta_D + \left( E_{nl} - V(r) \right)^2 - \left( M + S(r) \right)^2 \right] \psi_{nlm}(r, \Omega_D) = 0, \tag{16}$$

with

$$\Delta_D = \nabla_D^2 = \frac{1}{r^{D-1}} \frac{\partial}{\partial r} \left( r^{D-1} \frac{\partial}{\partial r} \right) - \frac{\Lambda_D^2(\Omega_D)}{r^2}, \tag{17}$$

where $E_{nl}$, $\Lambda_D^2(\Omega_D)/r^2$ and $\Omega_D$ are the energy eigenvalues, generation of the centrifugal barrier for $D$-dimensional space and the angular coordinates, respectively [45]. The eigenvalues of the $\Lambda_D^2(\Omega_D)$ are given by [45]

$$\Lambda_D^2(\Omega_D)Y_l^m(\Omega_D) = \frac{(D+2l-2)^2-1}{4}Y_l^m(\Omega_D), \quad D>1 \tag{18}$$

where $Y_l^m(\Omega_D)$ is the hyperspherical harmonics. For $D=2$, we have $\Lambda_{D=2}^2(\Omega_{D=2})Y_l^m(\Omega_{D=2}) = (m^2-1/4)Y_l^m(\Omega_{D=2})$ and also, For $D=3$, we have a familiar form $\Lambda_{D=3}^2(\Omega_{D=3})Y_l^m(\Omega_{D=3}) = l(l+1)Y_l^m(\Omega_{D=3})$. Using the procedure of separation of variables by means of the wave function $\psi_{nlm}(r,\Omega_D) = r^{-(D-1)/2}R_{nl}(r)Y_l^m(\Omega_D)$, Eq. (16) reduces to

$$\left[\frac{d^2}{dr^2} - (M^2-E_{nl}^2) - 2(E_{nl}V(r)+MS(r)) + V^2(r) - S^2(r) - \frac{(D+2l-2)^2-1}{4r^2}\right]R_{nl}(r) = 0. \tag{19}$$

Substituting the scalar and vector Yukawa potentials into Eq. (19), one obtains

$$\left[\frac{d^2}{dr^2} - \varepsilon^2 + \frac{(V_0^2-S_0^2)}{r^2}e^{-2ar} + \frac{2(MS_0+E_{nl}V_0)}{r}e^{-ar} - \frac{(D+2l-2)^2-1}{4r^2}\right]R_{nl}(r) = 0, \tag{20}$$

where $\varepsilon^2 = M^2 - E_{nl}^2$. We investigate the asymptotic behavior of $R_{nl}(r)$. First inspection of Eq. (20) shows that when $r$ approaches $\infty$, the asymptotic solution $R_0(r)$ of Eq. (20) satisfies the differential equation $\frac{d^2R_0(r)}{dr^2} - \varepsilon^2 R_0(r) = 0$, which assumes the solution of $R_0(r) \sim a_1 e^{-\varepsilon r}$, $a_1$ is a constant, as acceptable physical solution since the solution becomes finite as $r\to\infty$. Meanwhile as $r\to 0$, the asymptotic solution $R_\infty(r)$ of Eq. (20) satisfies the differential equation, $\frac{d^2R_\infty(r)}{dr^2} - \frac{(k^2-1/4)}{r^2}R_\infty(r) = 0$, where $k = \frac{1}{2}\sqrt{(D+2l-2)^2 + 4(S_0^2 - V_0^2)}$, which assumes the solution of $R_\infty(r) \sim a_2 r^{k+1/2} + a_3 r^{-(k+1/2)}$, where $a_2$ and $a_3$ are two constants. The term $r^{-(k+1/2)}$ is not satisfactory solution because it becomes infinite as $r\to 0$, but the term $r^{k+1/2}$ is well-behaved. Consequently, this asymptotic behavior of $R(r)$ suggests we choose the following appropriate ansatz for the radial wave function $R(r) = Ar^{k+1/2}e^{-\varepsilon r}F(r)$, where $F(r)$ is a hypergeometric function to be found from Eq. (20) in the region $r\in(0,\infty)$.

It is obvious that Eq. (20) does not admit an exact solution due to the presence of the singular terms $1/r$ and $1/r^2$. So we resort to approximation schemes to make approximations for these two terms [36-40] as

$$\frac{1}{r^2} \approx 4a^2 \frac{e^{-2ar}}{(1-e^{-2ar})^2}, \tag{21a}$$

$$\frac{1}{r} \approx 2a \frac{e^{-ar}}{(1-e^{-2ar})}, \tag{21b}$$

which are valid when $ar \ll 1$ [54-58]. Thus, the vector and scalar Yukawa potential in (1a) and (1b) can be approximated as

$$V(r) = -2aV_0 \frac{e^{-2ar}}{\left(1-e^{-2ar}\right)}, \tag{22a}$$

$$S(r) = -2aS_0 \frac{e^{-2ar}}{\left(1-e^{-2ar}\right)}, \tag{22b}$$

respectively. To show the accuracy of our approximation scheme, we plot the Yukawa potential (1a) and its approximation (22a) with parameter values $V_0 = \sqrt{2}$ , $a = 0.05V_0$ [59] in Figure 1. Now, substituting the approximations in (21) into (19), we can obtain

$$\left\{ \frac{d^2}{dr^2} - \varepsilon^2 + 4a^2(V_0^2 - S_0^2) \frac{e^{-4ar}}{(1-e^{-2ar})^2} \right.$$
$$\left. + 4a(MS_0 + E_{n,l}V_0) \frac{e^{-2ar}}{(1-e^{-2ar})} - a^2\left(D+2l-1\right)\left(D+2l-3\right) \frac{e^{-2ar}}{(1-e^{-2ar})^2} \right\} R_{nl}(r) = 0. \tag{23}$$

Making a suitable change of variables as $s = e^{-2ar}$, mapping $r \in (0,\infty)$ to $s \in (0,1)$, we can thereby recast Eq. (23) as follows

$$\frac{d^2R_{n,l}(s)}{ds^2} + \frac{1-s}{s(1-s)} \frac{dR_{n,l}(s)}{ds} + \frac{1}{s^2\left(1-s\right)^2} \left[ -\frac{\varepsilon^2}{4a^2}(1-s)^2 + (V_0^2 - S_0^2)s^2 \right.$$
$$\left. + \frac{(MS_0 + E_{nl}V_0)}{a} s(1-s) - \frac{\left(D+2l-2\right)^2 - 1}{4} s \right] R_{nl}(s) = 0. \tag{24}$$

The solution of Eq. (24) can be found by comparing it with Eq. (3) to get

$$c_1 = 1, \qquad p_0 = \frac{\varepsilon^2}{4a^2},$$

$$c_2 = 1, \qquad p_1 = \frac{2\varepsilon^2}{4a^2} + \frac{(MS_0 + E_{nl}V_0)}{a} - \frac{\left(D+2l-2\right)^2 - 1}{4}$$

$$c_3 = 1, \qquad p_2 = \frac{\varepsilon^2}{4a^2} - (V_0^2 - S_0^2) + \frac{(MS_0 + E_{nl}V_0)}{a}. \tag{25}$$

In addition, the values of constant coefficients $c_i$ $(i = 4,5,...,13)$ are found from Eq. (8) and displayed in table 1. Thus, using Eq. (6), the energy eigenvalue equation can be obtained as follows

$$\left(2n+1+\sqrt{(D+2l-2)^2-4(V_0^2-S_0^2)}+\frac{1}{a}\sqrt{M^2-E_{nl}^2}\right)^2$$
$$=\frac{M^2-E_{nl}^2}{a^2}+\frac{4(MS_0+E_{nl}V_0)}{a}+4\left(S_0^2-V_0^2\right), \tag{26a}$$

or it can be simply expressed as

$$\left(2n+1+\sqrt{(D+2l-2)^2+4\left(S_0^2-V_0^2\right)}+\frac{1}{a}\sqrt{M^2-E_{nl}^2}\right)^2=-\left(\frac{E_{nl}}{a}-2V_0\right)^2+\left(\frac{M}{a}+2S_0\right)^2, \tag{26b}$$

where $-M < E_{nl} < M$. Some approximated numerical results are calculated in Tables 2 to 4. The potential parameters are taken as $a = 0.05\,fm$ and $M = 1.0\,fm^{-1}$ [56]. In Table 2, we calculate the energy states for various values of dimensions $D$ and quantum numbers $n$ and $l$ when vector potential is different from scalar one, i.e. $V(r) \neq S(r)$. In Table 3, we obtain the energy levels for various values of dimensions $D$ and quantum numbers $n$ and $l$ when vector potential is equal to scalar one, i.e. $V(r) = S(r)$. Finally, in Table 4, we obtain the bound state energy for various values of dimensions $D$ and quantum numbers $n$ and $l$ when vector potential is opposite to scalar one, i.e. $V(r) = -S(r)$. As one can see from these Tables, there is an inter-dimensional degeneracy of eigenvalues of upper/lower dimensional system from the well-known eigenvalues of a lower/upper dimensional system by means of the transformation $(n,l,D) \to (n,l\pm1,D\mp2)$ [60]. Further, it is noticed that the energy of the state becomes less attractive with the increasing of the quantum numbers $n$, $l$ and the dimensional space $D$. In the case when $S_0 = \pm V_0$ (i.e., $\beta = \pm 1$), the energy levels: $E_{21} = E_{30}$, for any arbitrary dimension $D = 2-6$. However, when $\beta \neq \pm 1$ the energy levels $E_{21} \neq E_{30}$, that is, the degeneracy is removed. On the other hand, to show the accuracy of our approximation scheme, we have calculated the exact numerical energy eigenvalues of Eq. (20) without making approximation to the centrifugal term by using the AP method with arbitrary radial and orbital quantum numbers and dimension $n$, $l$ and $D$. We have found the percentage error $\left|\frac{E_{\text{approx}}-E_{\text{exact}}}{E_{\text{exact}}}\right| \times 100\%$ for a few states in table 2 as 2.7403%, 2.4987 % and 1.8994 % for $(n,l,D)=(3,1,2)$; $(3,1,3)$ and $(3,1,6)$; respectively. It is found that the exact numerical energy states obtained by means of Eq. (20) are in good agreement with the approximate ones obtained by Eq. (24) for low screening regime, i.e., $ar \ll 1$ since our approximation works well only for the lowest energy states [56].

In table 2, when $n=1$ and $l=0$, all the energy states are attractive (negative) for $2 \leq D \leq 5$ and become less attractive with increasing dimension. However, when $n \geq 2$ and $l \geq 0$, all energy states become more repulsive (positive) with the increasing dimension. In table 3, for $n=1$ and $l=0$, the system is attractive for all states when $D \geq 2$ and becomes less attractive when dimension increases. When $n=2$ and $l=0$, the system is less attractive for all states when $2 \leq D \leq 5$ and repulsive for $D>5$. When $n=2$ and $l=1$, together with $n=3$ and $l=0$, the system is less attractive when $D=2,3$ and more repulsive for $D>3$ when dimension increases. Further when $n \geq 3$ and $l \geq 1$, its more repulsive for $D \geq 2$ as the dimension increases. In Table 3, when $n=1$ and $l=0$, all the energy states are less repulsive with increasing dimension. However, when $n=2$ and $l=0$, the system is less repulsive for $2 \leq D \leq 5$ and more attractive for $D>5$. When $n=2$ and $l=1$, together with $n=3$ and $l=0$, the system is less repulsive when $D=2,3$ and more attractive for $D>3$ for all states when dimension increases. Finally, when $n \geq 3$ and $l \geq 1$, its more attractive for $D \geq 2$ as the dimension increases.

Let us now calculate the corresponding eigenfunctions, we use the relations in (7) to obtain the necessary functions

$$\rho(s)=s^{\frac{\varepsilon}{a}}(1-s)^{\sqrt{4(S_0^2-V_0^2)+(D+2l-2)^2}},$$

$$\phi(s)=s^{\frac{\varepsilon}{2a}}(1-s)^{\frac{1}{2}\left(-1+\sqrt{4(S_0^2-V_0^2)+(D+2l-2)^2}\right)},$$

$$y_n(s)=P_n^{\left(\frac{\varepsilon}{a},\sqrt{4(S_0^2-V_0^2)+(D+2l-2)^2}\right)}(1-2s),$$

$$R_{nl}(s)=N_{nl}s^{\frac{\varepsilon}{2a}}(1-s)^{\frac{1}{2}\left(-1+\sqrt{4(S_0^2-V_0^2)+(D+2l-2)^2}\right)}P_n^{\left(\frac{\varepsilon}{a},\sqrt{4(S_0^2-V_0^2)+(D+2l-2)^2}\right)}(1-2s), \quad (27)$$

where $N_{nl}$ is the normalization constant. The radial wave functions can also be rewritten in a more convenient form in terms of the potential parameters as

$$R_{nl}(r)=N_{nl}e^{\varepsilon r}\left(1-e^{-2ar}\right)^{\frac{1}{2}\left(-1+\sqrt{4(S_0^2-V_0^2)+(D+2l-2)^2}\right)}$$

$$\times P_n^{\left(\frac{\varepsilon}{a},\sqrt{4(S_0^2-V_0^2)+(D+2l-2)^2}\right)}\left(1-2e^{-2ar}\right), \tag{28}$$

which well-behaved at boundaries, i.e., $r=0$ and $r\to\infty$.

From Eq. (26b), we obtain the energy eigenvalues of the KG particles in 2- and 3-dimensional spaces (2D) and (3D), respectively, as

$$\left(2n+1+2\sqrt{\left(m-\frac{1}{2}\right)^2+\left(S_0^2-V_0^2\right)}+\frac{1}{a}\sqrt{M^2-E_{nm}^2}\right)^2=-\left(\frac{E_{nm}}{a}-2V_0\right)^2+\left(\frac{M}{a}+2S_0\right)^2, \tag{29a}$$

$$\left(2n+1+2\sqrt{\left(l+\frac{1}{2}\right)^2+\left(S_0^2-V_0^2\right)}+\frac{1}{a}\sqrt{M^2-E_{nl}^2}\right)^2=-\left(\frac{E_{nl}}{a}-2V_0\right)^2+\left(\frac{M}{a}+2S_0\right)^2. \tag{29b}$$

where, in (29a), we inserted $l\to m-1/2$ in 2D.

When we set $V(r)\to V(r)/2,\ S(r)\to S(r)/2,\ E_{nl}+M\to 2\mu/\hbar^2$ and $E_{nl}-M\to E_{nl}$ [56], we can obtain the solution in the non-relativistic limit of the Yukawa problem. Here $\mu=m_1m_2/(m_1+m_2)$ is the reduced mass with $m_1$ and $m_2$ stand for masses of the electron $e$ and the atom $Ze$, respectively. Under these conditions, one can obtain the non-relativistic energy eigenvalues of the Yukawa potential as [40]

$$E_{nl}=-\frac{\hbar^2}{2\mu}\left[\left(n+\frac{D}{2}+l-\frac{1}{2}\right)a-\frac{\mu V_0}{\hbar^2\left(n+\frac{D}{2}+l-\frac{1}{2}\right)}\right]^2, \tag{30}$$

and the corresponding radial wave functions reduce to

$$R_{nl}(r)=N_{n,l}e^{-\frac{1}{\hbar}\sqrt{2\mu E_{nl}}r}(1-e^{-2ar})^{l+1}P_n^{\left(\frac{1}{\hbar a}\sqrt{2\mu E_{nl}},2l+1-\frac{1}{\hbar a}\sqrt{2\mu E_{nl}}\right)}(1-2e^{-2ar}). \tag{31}$$

Also, when the screening parameter $a$ approaches zero, the potential (1) reduces to a Coulomb potential. Thus, in this limit the energy eigenvalues of (30) become the energy levels of the Coulomb interaction, i.e.,

$$E_{nl}=-\frac{\mu V_0^2}{2\hbar^2\left(n+\frac{D}{2}+l-\frac{1}{2}\right)^2}, \tag{32}$$

which is identical to Refs. [45, 58] when $D=3$.

## 4. Concluding Remarks

We have used a simple shortcut of the NU method as well as an appropriate approximation for the strong and soft singular terms to obtain approximate analytical bound states solutions of the $D$-dimensional KG equation for scalar and vector Yukawa potentials. Numerical tests using energy calculations show the existence of inter-dimensional degeneracy of energy states prevailing to the following transformation: $(n,l,D)\rightarrow(n,l\pm1,D\mp2)$ as shown in Tables 2 to 4. Further, it is noted that when $V_0\neq S_0,$ the weakly attractive system turns to become less attractive with the increasing dimension $D$ and afterwards to strongly repulsive with the increasing quantum numbers $n$ and $l.$ In the case when $V_0=S_0,$ the strongly attractive system becomes less attractive with the increasing $D$ and to strongly repulsive with the increasing $n$ and $l.$ However, when $V_0=-S_0,$ the strongly repulsive system becomes less repulsive with the increasing $D$ and strongly attractive with the increasing of $n$ and $l.$ We have also calculated the exact numerical energy eigenvalues of Eq. (20) via the numerical AP method. Our results show the agreement between the approximate solution of Eq. (24) and the exact solution of Eq. (20).. In the present solution, we have not faced any cumbersome and time-consuming procedures in obtaining the numerical and analytical eigenvalues and wave functions of the problem by a simple methodology. The results are applicable to a wide range of relevant fields.

**Acknowledgments**

We thank the kind referee for the positive suggestions and critics which have greatly improved the present paper. S.M. Ikhdair acknowledges the partial support of the Scientific and Technological Research Council of Turkey. M. Hamzavi thanks his host Institution, KTH-Mechanics, Royal Institute of technology, S-100 44 Stockholm, Sweden.

**Table 1.** The values of the parametric constants used to calculate the energy eigenvalues and eigenfunctions

| constant | Analytical value |
|---|---|
| $c_4$ | $0$ |
| $c_5$ | $-\frac{1}{2}$ |
| $c_6$ | $\frac{1}{4}+\frac{\varepsilon^2}{4a^2}-(V_0^2+S_0^2)+\frac{(MS_0+E_{n,l}V_0)}{a}$ |
| $c_7$ | $-\frac{2\varepsilon^2}{4a^2}-\frac{(MS_0+E_{n,l}V_0)}{a}+\frac{(D+2l-1)(D+2l-3)}{4}$ |
| $c_8$ | $\frac{\varepsilon^2}{4a^2}$ |
| $c_9$ | $-(V_0^2+S_0^2)+\frac{(D+2l-1)(D+2l-3)+1}{4}$ |
| $c_{10}$ | $2\sqrt{\frac{\varepsilon^2}{4a^2}}$ |
| $c_{11}$ | $2\sqrt{-(V_0^2+S_0^2)+\frac{(D+2l-1)(D+2l-3)+1}{4}}$ |
| $c_{12}$ | $\sqrt{\frac{\varepsilon^2}{4a^2}}$ |
| $c_{13}$ | $-\frac{1}{2}+\sqrt{-(V_0^2+S_0^2)+\frac{(D+2l-1)(D+2l-3)+1}{4}}$ |

**Table 2.** The energy levels of the KG particles in the field of scalar and vector Yukawa potentials for various $D$ , $n$ and $l$ values with $V_0 = 4.5$ and $S_0 = 5$.

| | $E_{n,l}(fm^{-1})$ | | | | | | | | | |
|---|---|---|---|---|---|---|---|---|---|---|
| $n,l$<br>D | 1,0<br>NU | 1,0<br>AP | 2,0<br>NU | 2,0<br>AP | 2,1<br>NU | 2,1<br>AP | 3,0<br>NU | 3,0<br>AP | 3,1<br>NU | 3,1<br>AP |
| **2** | -0.214913 | -0.219440 | 0.112670 | 0.105122 | 0.174406 | 0.166646 | 0.368030 | 0.357228 | 0.415044 | 0.403974 |
| **3** | -0.194399 | -0.198948 | 0.129002 | 0.121398 | 0.240311 | 0.232333 | 0.380497 | 0.369622 | 0.464924 | 0.453590 |
| **4** | -0.137038 | -0.141652 | 0.174406 | 0.166646 | 0.317325 | 0.309115 | 0.415044 | 0.403974 | 0.522848 | 0.511245 |
| **5** | -0.052932 | -0.057646 | 0.240311 | 0.232333 | 0.397892 | 0.389478 | 0.464924 | 0.453590 | 0.583068 | 0.571249 |
| **6** | 0.046537 | 0.041712 | 0.317325 | 0.309115 | 0.476916 | 0.468366 | 0.522848 | 0.511245 | 0.641793 | 0.629861 |

**Table 3.** Bound state energy levels of the KG particle subjects to scalar and vector Yukawa potentials for various $D$ , $n$ and $l$ values with $V_0 = S_0 = 5$ .

| | $E_{n,l}(fm^{-1})$ | | | | | | | | | |
|---|---|---|---|---|---|---|---|---|---|---|
| $n,l$<br>D | 1,0<br>NU | 1,0<br>AP | 2,0<br>NU | 2,0<br>AP | 2,1<br>NU | 2,1<br>AP | 3,0<br>NU | 3,0<br>AP | 3,1<br>NU | 3,1<br>AP |
| **2** | -0.795853 | -0.784988 | -0.506330 | -0.493815 | -0.190290 | -0.194878 | -0.190290 | -0.194878 | 0.098154 | 0.090716 |
| **3** | -0.659219 | -0.660899 | -0.347361 | -0.351045 | -0.040662 | -0.046067 | -0.040662 | -0.046067 | 0.224479 | 0.216023 |
| **4** | -0.506330 | -0.508584 | -0.190290 | -0.194878 | 0.098154 | 0.092028 | 0.098154 | 0.092028 | 0.337828 | 0.328501 |
| **5** | -0.347361 | -0.350140 | -0.040662 | -0.046067 | 0.224479 | 0.217748 | 0.224479 | 0.217748 | 0.438486 | 0.428448 |
| **6** | -0.190290 | -0.193533 | 0.098154 | 0.092028 | 0.337828 | 0.330622 | 0.337828 | 0.330622 | 0.527179 | 0.516612 |

**Table 4.** Bound state energy levels of the KG particle in the field of scalar and vector Yukawa potentials for various $D$ , $n$ and $l$ values with $V_0 = -5$ and $S_0 = 5$.

| | $E_{n,l}(fm^{-1})$ | | | | | | | | | |
|---|---|---|---|---|---|---|---|---|---|---|
| $n,l$ | 1,0 | 1,0 | 2,0 | 2,0 | 2,1 | 2,1 | 3,0 | 3,0 | 3,1 | 3,1 |
| D | NU | AP | NU | AP | NU | AP | NU | AP | NU | AP |
| **2** | 0.795853 | 0.784988 | 0.506330 | 0.493814 | 0.190290 | 0.194878 | 0.190290 | 0.194878 | -0.098154 | -0.090716 |
| **3** | 0.659219 | 0.660899 | 0.347361 | 0.351046 | 0.040661 | 0.046069 | 0.040661 | 0.046069 | -0.224478 | -0.216023 |
| **4** | 0.506330 | 0.508584 | 0.190290 | 0.194878 | -0.098154 | -0.092028 | -0.098154 | -0.092028 | -0.337828 | -0.328501 |
| **5** | 0.347361 | 0.350140 | 0.040661 | 0.046069 | -0.224479 | -0.217748 | -0.224479 | -0.217748 | -0.438485 | -0.428448 |
| **6** | 0.190290 | 0.193533 | -0.098154 | -0.092028 | -0.337828 | -0.330621 | -0.337828 | -0.330621 | -0.527180 | -0.516612 |

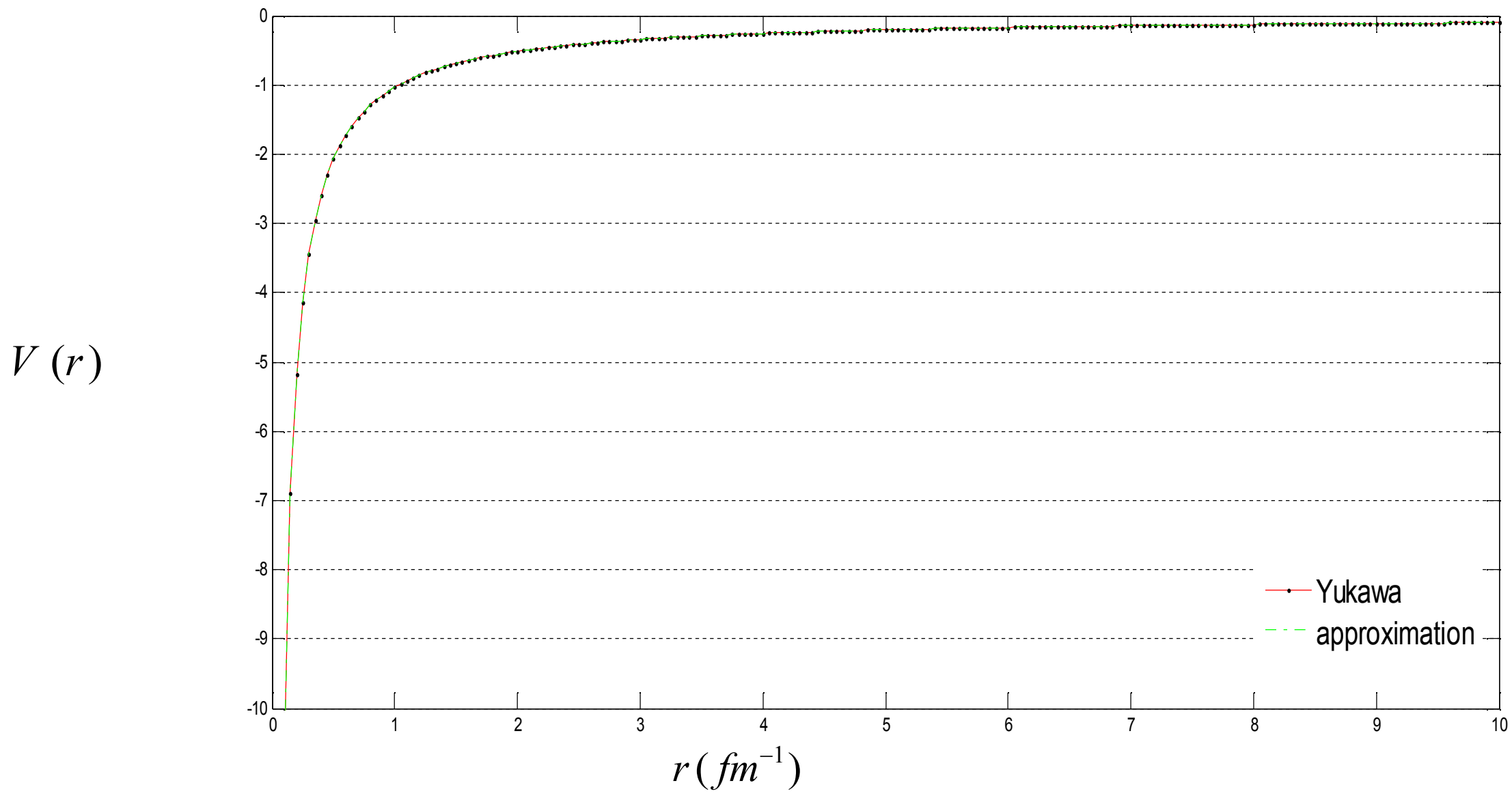

**Figure 1:** The Yukawa potential (red line) and its approximation in Eq. (16) (green line).